\documentclass[preprint,12pt]{elsarticle}

\usepackage{amssymb}

\usepackage{amsmath,amsthm,amssymb,mathrsfs}
\usepackage{bm}

\journal{Journal of Magnetism and Magnetic Materials}

\begin{document}

\begin{frontmatter}




\title{An analytical computation of magnetic field generated from a cylinder ferromagnet}


\author{Tomohiro Taniguchi 
}


\address{
 National Institute of Advanced Industrial Science and Technology (AIST), Spintronics Research Center, Tsukuba, Ibaraki 305-8568, Japan, 
}



\begin{abstract}
An analytical formulation to compute a magnetic field generated from an uniformly magnetized cylinder ferromagnet is developed. 
Exact solutions of the magnetic field generated from the magnetization pointing in an arbitrary direction are derived, 
which are applicable both inside and outside the ferromagnet. 
The validities of the present formulas are confirmed by comparing them with demagnetization coefficients estimated in earlier works. 
The results will be useful for designing practical applications, such as high-density magnetic recording and microwave generators, 
where nanostructured ferromagnets are coupled to each other through the dipole interactions 
and show cooperative phenomena such as synchronization. 
As an example, the magnetic field generated from a spin torque oscillator 
for magnetic recording based on microwave assisted magnetization reversal is studied. 
\end{abstract}


\end{frontmatter}





\section{Introduction}
\label{sec:Introduction}

The computation of magnetic field generated from a ferromagnetic body has been an important problem 
in the field of fundamental magnetism, mathematical physics, and applied physics 
\cite{osborn45,joseph65,joseph66,druyvesteyn71,chen91,aharoni98,bertotti98,jackson98,guslienko99,metlov00,aharoni01,tandon04,beleggia06,hubert08}. 
In particular, an analytical, as well as numerical, calculation of the demagnetization coefficient (or tensor) has been a main topic in these earlier works 
because it determines the magnetization configuration in equilibrium state. 
For example, Ref. \cite{joseph66} studies the magnetometric and ballistic demagnetization factors of a cylinder ferromagnet, 
where the former is obtained by averaging the magnetic field inside the ferromagnet over the volume, 
whereas the latter is calculated from the average over the middle cross section. 
The stray fields from a perpendicularly magnetized and an in-plane magnetized cylinder ferromagnets in a periodic structure were calculated in Refs. \cite{druyvesteyn71,guslienko99,metlov00}. 


The problem to compute a magnetic field is still interesting for current magnetism because of recent developments in the fabrication technology of nanostructured ferromagnets. 
For example, material investigations of magnetic multilayers revealed that 
magnetization in a nanostructured ferromagnet has a capability of pointing to the direction perpendicular to the film-plane \cite{yakata09,ikeda10,kubota12} 
due to a magnetic anisotropy induced at CoFeB/MgO interface, 
although a large demagnetization field along the perpendicular direction prefers an in-plane magnetized structure. 
It was also shown that the magnetization dynamics, such as switching and auto-oscillation, can be excited by spin transfer effect \cite{slonczewski96} 
in nanostructured ferromagnetic multilayers \cite{kiselev03}. 
In these dynamic states, several ferromagnets located close to each other within a nanoscale distance affect each other 
and show a coupled motion of the magnetizations, such as a resonant switching and synchronized oscillation, through stray (dipole) fields from each ferromagnet 
\cite{zhu08,suto12,belanvosky12,belanvosky13,kudo14,suto14,locatelli15,qu15,araujo15,kudo15,chen16,suto16PRA,suto16}. 
Such magnetization dynamics and/or cooperative phenomena are useful for practical applications 
such as high-density magnetic recording, microwave generators, and neuromorphic architectures. 
Therefore, it will be highly desirable for designing these devices to develop a theoretical method to compute the magnetic field generated from nanostructured ferromagnets. 
It was pointed out that a point dipole model is not sufficient to evaluate the stray field due to finite size effect \cite{chen16}. 
Micromagnetic simulation is one solution to overcome this problem \cite{brown65,labonte69,hayashi71,schryer74,shir78,torre85,mansuripur88,nakatani89}. 
At the same time, development of an analytical approach to evaluate the stray field will also be useful 
from the perspective of giving a clue to the relation between the field and physical parameters.




Motivated by these works, in this paper, we develop analytical formulas of magnetic field generated from a cylinder ferromagnet. 
The magnetic fields are calculated for both perpendicularly and in-plane magnetized cases. 
The magnetic field generated from the magnetization pointing in an arbitrary direction 
can also be obtained as a linear combination of these cases. 
Therefore, the formulas can be applied to not only the static (equilibrium) state of the magnetization 
but also to the dynamical state excited by, for example, microwaves and spin-transfer torque, 
where the magnetization direction is neither in-plane nor perpendicular direction. 
It should also be noted that the derived formulas are applicable in the whole space both inside and outside the ferromagnet. 
The validity of the present calculations is confirmed by comparing the demagnetization coefficient derived from our formula 
with that found in earlier works. 
As mentioned above, many efforts have been made to compute the magnetic field from a cylinder ferromagnet. 
The present study provides a simplified formula of the magnetic fields in terms of the complete elliptic integrals. 
The insight brought forward in our analytical derivation provides simple and lucid formulation of the continuity/discontinuity of the magnetic fields at the surface of the magnet 
which plays an important role in the stray field distribution in practical devices. 
The resultant formulation of the boundary condition is perfectly consistent with the classical electromagnetic theory. 
As an example of the application of the present results, 
the generation of a magnetic field from a spin torque oscillator 
for magnetic recording based on microwave assisted magnetization reversal is also studied. 


The paper is organized as follows. 
In Sec. \ref{sec:System description}, we describe the system we adopted in the study. 
The magnetic field generated from a perpendicularly magnetized cylinder ferromagnet is calculated in Sec. \ref{sec:Magnetic field generated from a perpendicularly magnetized ferromagnet}. 
On the other hand, Sec. \ref{sec:Magnetic field generated from an in-plane magnetized ferromagnet} shows 
the calculation of the field generated from an in-plane magnetized cylinder ferromagnet. 
In Sec. \ref{sec:Application}, we show an example of the calculation of the magnetic field distribution from a spin torque oscillator. 
The summary of this work is discussed in Sec. \ref{sec:Summary}. 




\begin{figure}
\centerline{\includegraphics[width=1.0\columnwidth]{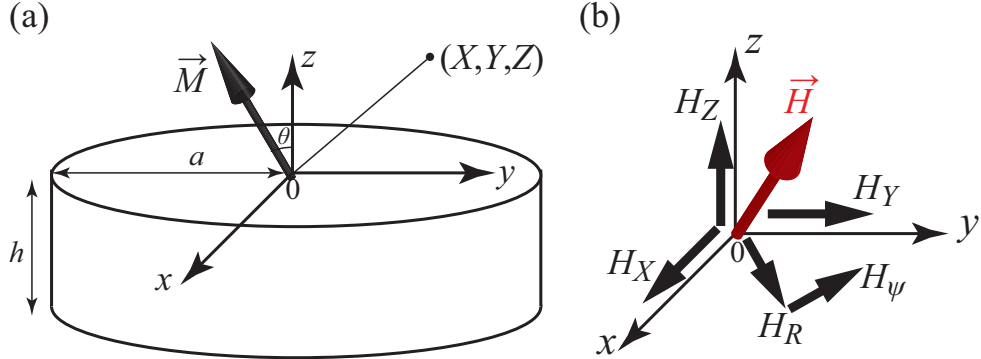}}\vspace{-3.0ex}
\caption{
        (a) Schematic illustration of system under consideration. 
            The radius and thickness of the ferromagnet are $a$ and $h$, respectively. 
        (b) The components of the magnetic field $\vec{H}$ in the Cartesian and cylindrical coordinates. 
         \vspace{-3ex}}
\label{fig:fig1}
\end{figure}




\section{System description}
\label{sec:System description}

The system used in this study is schematically shown in Fig. \ref{fig:fig1}(a). 
A cylinder ferromagnet with the radius $a$ and thickness $h$ is placed in a coordinate 
where the origin locates at the center of the top surface. 
Note that the cylinder-shaped ferromagnet is a common structure widely used in the nanostructured ferromagnetic device 
because its axial symmetry makes it relatively easy to control magnetization dynamics, such as switching and auto-oscillation. 
It has been clarified that the magnetization dynamics in such ferromagnets is well described by macrospin mode, 
i.e., assuming an uniformly magnetized ferromagnet gives reasonable results showing good agreement with experiments \cite{grollier03}. 
For convention, we assume that the magnetization $\vec{M}$ ($M=|\vec{M}|$) lies in the $xz$ plane 
with the tilted angle $\theta$ from the $z$ axis. 
We denote a point measuring the magnetic field generated by the ferromagnet as $(X,Y,Z)$. 
Note that point $(X,Y,Z)$ can be either inside or outside of the ferromagnet. 


We denote the magnetic field generated by a perpendicularly magnetized ferromagnet ($\vec{M}=M \vec{e}_{z}$) as $\vec{H}_{Z}$. 
Similarly, the magnetic field generated from an in-plane magnetized ferromagnet ($\vec{M}=M \vec{e}_{x}$) is denoted as $\vec{H}_{X}$. 
Then, the magnetic field $\vec{H}$ 
generated from a magnetization $\vec{M}= M (\cos\theta \vec{e}_{z} + \sin\theta \vec{e}_{x}$) pointing to an arbitrary direction is given by 
\begin{equation}
  \vec{H}
  =
  \vec{H}_{Z}
  \cos\theta
  +
  \vec{H}_{X}
  \sin\theta.
  \label{eq:field_def}
\end{equation}
In the following, we denote the $j$ component of $\vec{H}_{i}$ ($i=X,Z$) as $H_{ij}$, 
where the subscript $j$ stands for $j=X,Y,Z$ in the Cartesian coordinate.
Note that the cylindrical coordinate will be used in the following calculations. 
In such case, the subscript $j$ stands for $j=R,\psi,Z$ in the cylindrical coordinate, 
where $R$ and $\psi$ are the radius and azimuth angle satisfying 
$X=R \cos\psi$ and $Y=R \sin\psi$. 
The components of $\vec{H}_{i}$ in both the Cartesian and cylindrical coordinates are schematically shown in Fig. \ref{fig:fig1}(b). 
They satisfy 
\begin{equation}
  H_{iX}
  =
  H_{iR} 
  \cos\psi 
  - 
  H_{i\psi} 
  \sin\psi,
  \label{eq:H_X_R_psi}
\end{equation}
\begin{equation}
  H_{iY}
  =
  H_{iR} 
  \sin\psi 
  + 
  H_{i\psi} 
  \cos\psi.
  \label{eq:H_Z_R_psi}
\end{equation}


In the following sections, the solutions of $\vec{H}_{Z}$ and $\vec{H}_{X}$ will be given. 
Since the calculations are rather complex, we point out beforehand the index numbers of key equations obtained in this study. 
The components of $\vec{H}_{Z}$ in the cylindrical coordinate are Eqs. (\ref{eq:H_Z_R}) and (\ref{eq:H_Z_Z}).
The components of $\vec{H}_{X}$ are given by 
Eqs. (\ref{eq:H_X_R}), (\ref{eq:H_X_Z}), and (\ref{eq:H_X_psi}) for $Z>0$ and $Z<-h$, 
whereas those for $-h<Z<0$ are given by Eqs. (\ref{eq:H_X_R_inside}), (\ref{eq:H_X_Z_inside}), and (\ref{eq:H_X_psi_inside}). 


\section{Magnetic field generated from a perpendicularly magnetized ferromagnet}
\label{sec:Magnetic field generated from a perpendicularly magnetized ferromagnet}

In this section, we focus on $\vec{H}_{Z}$, which is a magnetic field generated from a perpendicularly magnetized cylinder ferromagnet. 


\subsection{Magnetic potential}

The magnetic field is calculated as a gradient of the magnetic potential. 
In the present case, the magnetic poles appeared on the surfaces at $z=-h$ and $z=0$ contribute to the potential, 
which is given by 
\begin{equation}
\begin{split}
  V_{\rm p}
  =&
  M 
  \int_{0}^{a}
  d \rho 
  \int_{0}^{2\pi}
  \rho d \phi
  \frac{1}{\sqrt{(X-\rho \cos\phi)^{2} + (Y-\rho \sin\phi)^{2} + Z^{2}}}
\\
  &-
  M
  \int_{0}^{a}
  d \rho 
  \int_{0}^{2\pi}
  \rho d \phi
  \frac{1}{\sqrt{(X-\rho \cos\phi)^{2} + (Y-\rho \sin\phi)^{2} + (Z+h)^{2}}}. 
  \label{eq:V_p}  
\end{split}
\end{equation}
The Cartesian components of $\vec{H}_{Z}$ can be obtained as the gradient of $V_{\rm p}$, 
i.e., $\vec{H}_{Z}=(H_{ZX},H_{ZY},H_{ZZ})=(-\partial V_{\rm p}/\partial X, -\partial V_{\rm p}/\partial Y, -\partial V_{\rm p}/\partial Z)$. 
For the current calculations, however, it is useful to derive the components of $\vec{H}_{Z}$ in the cylindrical coordinate system, as mentioned earlier. 

Due to the axial symmetry of the present case, $H_{Z \psi}=0$. 
To calculate $H_{ZR}$ and $H_{ZZ}$, on the other hand, 
it is convenient to express the integrands of Eq. (\ref{eq:V_p}) as 
\begin{equation}
\begin{split}
  &
  \frac{\rho}{\sqrt{(X-\rho \cos\phi)^{2} + (Y-\rho \sin\phi)^{2} + Z^{2}}}
  -
  \frac{\rho}{\sqrt{(X-\rho \cos\phi)^{2} + (Y-\rho \sin\phi)^{2} + (Z+h)^{2}}}
\\
  &=
  \frac{\rho}{\sqrt{r^{2}+Z^{2}}}
  -
  \frac{\rho}{\sqrt{r^{2}+(Z+h)^{2}}},
\end{split}
\end{equation}
where $r^{2}=R^{2}+\rho^{2}-2 R \rho \cos(\phi-\psi)$. 
Using the Gegenbauer's addition theorem for the Bessel function $J_{n}(x)$ \cite{watson95}, 
\begin{equation}
\begin{split}
  \frac{\rho}{\sqrt{r^{2}+Z^{2}}}
  &=
  \rho 
  \int_{0}^{\infty} 
  d k 
  e^{-k|Z|}
  J_{0}(kr)
\\
  &=
  \rho 
  \int_{0}^{\infty}
  dk 
  e^{-k|Z|}
  \left[
    J_{0}(kR)
    J_{0}(k \rho)
    +
    2 \sum_{m=1}^{\infty}
    J_{m}(kR) 
    J_{m}(k \rho)
    \cos m \theta
  \right],
  \label{eq:Gegenbauer_formula}
\end{split}
\end{equation}
and the formula $\int d \rho [\rho J_{0}(k \rho)]=\rho J_{1}(k \rho)/k$, 
we notice that 
\begin{equation}
  V_{\rm p}
  =
  2\pi a M 
  \int_{0}^{\infty}
  d k 
  \frac{e^{-k|Z|} - e^{-k|Z+h|}}{k}
  J_{0}(kR)
  J_{1}(ka).
\end{equation}
There are similar examples of the magnetic potential involving the integrals of the Bessel functions, 
such as static \cite{guslienko01} and dynamical \cite{metlov13} magnetic vortices, and 
quasiuniformly magnetized cylinder ferromagnet \cite{metlov04}. 
In the present case, we have reduced the integral with the Bessel functions to a simplified form with the complete elliptic integrals, as shown below. 


\subsection{Magnetic field in radial direction}

The component of the magnetic field in the radial direction in the cylindrical coordinate, $H_{R}$, is obtained as 
$H_{ZR}=-\partial V_{\rm p}/\partial R$. 
Using the formula $\partial J_{0}(kR)/\partial R=-kJ_{1}(kR)$, it is found that \cite{joseph65} 
\begin{equation}
\begin{split}
  H_{ZR}
  &=
  2\pi a M 
  \int_{0}^{\infty}
  dk 
  \left(
    e^{-k|Z|}
    -
    e^{-k|Z+h|}
  \right)
  J_{1}(kR)
  J_{1}(ka)
\\
  &=
  4 \sqrt{\frac{a}{R}}
  M 
  \left\{
    \frac{1}{\kappa_{1}}
    \left[
      \left(
        1
        -
        \frac{\kappa_{1}^{2}}{2}
      \right)
      \mathsf{K}(\kappa_{1})
      -
      \mathsf{E}(\kappa_{1})
    \right]
    -
    \frac{1}{\kappa_{2}}
    \left[
      \left(
        1
        -
        \frac{\kappa_{2}^{2}}{2}
      \right)
      \mathsf{K}(\kappa_{2})
      -
      \mathsf{E}(\kappa_{2})
    \right]
  \right\},
  \label{eq:H_Z_R}
\end{split}
\end{equation}
where $\mathsf{K}(\kappa)$ and $\mathsf{E}(\kappa)$ are the first and second kinds of Jacobi's complete elliptic integrals 
with the modulus
\begin{equation}
  \kappa_{1}
  =
  \sqrt{
    \frac{4 aR}{Z^{2}+(a+R)^{2}}
  }.
  \label{eq:kappa_1}
\end{equation}
We use the integral formula of the Bessel function \cite{eason55,byrd13} in the derivation of Eq. (\ref{eq:H_Z_R}). 
The modulus $\kappa_{2}$ is obtained by replacing $Z$ in $\kappa_{1}$ with $Z+h$. 
We note that $H_{ZR}(X,Y,Z_{0})=-H_{ZR}(X,Y,-Z_{0}-h)$ is satisfied for a positive $Z_{0}$. 


\subsection{Magnetic field in perpendicular direction}

On the other hand, the field along the $z$ direction, $H_{ZZ}=-\partial V_{\rm p}/\partial Z$, is given by 
\begin{equation}
\begin{split}
  H_{ZZ}
  &=
  2\pi a M 
  \int_{0}^{\infty}
  dk 
  \left[
    {\rm sgn}(Z)
    e^{-k|Z|}
    -
    {\rm sgn}(Z+h)
    e^{-k|Z+h|}
  \right]
  J_{0}(kR)
  J_{1}(ka)
\\
  &=
  2\pi M 
  \left[
    {\rm sgn}(Z)
    U(Z)
    -
    {\rm sgn}(Z+h)
    U(Z+h)
  \right]
\\
  &=
  -4 M 
  \left[
    {\rm sgn}(Z)
    \mathsf{E}(\Lambda_{1})
    \mathsf{F}(\beta_{1},\sqrt{1-\Lambda_{1}^{2}})
    -
    {\rm sgn}(Z+h)
    \mathsf{E}(\Lambda_{2})
    \mathsf{F}(\beta_{2},\sqrt{1-\Lambda_{2}^{2}})
  \right.
\\
  &+
    {\rm sgn}(Z)
    \mathsf{K}(\Lambda_{1})
    \mathsf{E}(\beta_{1},\sqrt{1-\Lambda_{1}^{2}})
    -
    {\rm sgn}(Z+h)
    \mathsf{K}(\Lambda_{2})
    \mathsf{E}(\beta_{2},\sqrt{1-\Lambda_{2}^{2}})
\\
  &-
  \left.
    {\rm sgn}(Z)
    \mathsf{K}(\Lambda_{1})
    \mathsf{F}(\beta_{1},\sqrt{1-\Lambda_{1}^{2}})
    +
    {\rm sgn}(Z+h)
    \mathsf{K}(\Lambda_{2})
    \mathsf{F}(\beta_{2},\sqrt{1-\Lambda_{2}^{2}})
  \right]
\\
  &-
  4\pi M 
  \left[
    \Theta(-h) 
    -
    \Theta(0)
  \right],
  \label{eq:H_Z_Z}
\end{split}
\end{equation}
where $U(z)$ is defined as 
\begin{equation}
\begin{split}
  U(Z)
  &=
  a
  \int_{0}^{\infty}
  dk 
  e^{-k|Z|}
  J_{0}(kR)
  J_{1}(ka)
\\
  &=
  1 
  - 
  \frac{2}{\pi} 
  \left[ 
    \mathsf{E}(\Lambda_{1}) 
    \mathsf{F}(\beta_{1},\sqrt{1-\Lambda_{1}^{2}}) 
    + 
    \mathsf{K}(\Lambda_{1}) 
    \mathsf{E}(\beta_{1},\sqrt{1-\Lambda_{1}^{2}}) 
    - 
    \mathsf{K}(\Lambda_{1}) 
    \mathsf{F}(\beta_{1},\sqrt{1-\Lambda_{1}^{2}}) 
  \right],
  \label{eq:U_Z}
\end{split}
\end{equation}
whereas ${\rm sgn}(Z)$ and $\Theta(Z)$ are the sign and step functions, respectively. 
Note that the factor $\Theta(-h)-\Theta(0)$ is finite only inside the ferromagnet, $-h<Z<0$. 
The formula of the Laplace transformation of the Bessel function is used \cite{byrd13}. 
Here, $\mathsf{F}(\beta,\Lambda)$ and $\mathsf{E}(\beta,\Lambda)$ are the first and second kinds of Jacobi's incomplete elliptic integrals 
with the modulus $\Lambda$ and amplitude $\beta$ given by 
\begin{equation}
\begin{split}
  \Lambda_{1}^{2}
  =&
  \frac{a^{2}-Z^{2}-R^{2}+\sqrt{(a^{2}+Z^{2}+R^{2})^{2}-4a^{2}R^{2}}}{Z^{2}+R^{2}-a^{2}+\sqrt{(a^{2}+Z^{2}+R^{2})^{2}-4a^{2}R^{2}}}
\\
  &\times
  \frac{R^{2}-Z^{2}-a^{2}+\sqrt{(a^{2}+Z^{2}+R^{2})^{2}-4a^{2}R^{2}}}{Z^{2}-R^{2}+a^{2}+\sqrt{(a^{2}+Z^{2}+R^{2})^{2}-4a^{2}R^{2}}},
  \label{eq:Lambda_1}
\end{split}
\end{equation}
\begin{equation}
  \beta_{1}
  =
  \sin^{-1}
  \sqrt{
    \frac{1}{2}
    \left[
      1
      +
      \frac{Z^{2}+R^{2}-a^{2}}{\sqrt{(Z^{2}+R^{2}+a^{2})^{2}-4a^{2}R^{2}}}
    \right]
  }.
  \label{eq:beta_1}
\end{equation}
Note that $\Lambda_{2}$ and $\beta_{2}$ are obtained by replacing $Z$ in $\Lambda_{1}$ and $\beta_{1}$ with $Z+h$. 
We note that the sign functions are necessary in Eq. (\ref{eq:H_Z_Z}) to guarantee 
$H_{ZZ}(X,Y,Z_{0})=H_{ZZ}(X,Y,-Z_{0}-h)$ for a positive $Z_{0}$. 


\subsection{Short summary of this section}

Equations (\ref{eq:H_Z_R}) and (\ref{eq:H_Z_Z}) give analytical formulas of the magnetic field generating by a perpendicularly magnetized ferromagnet. 
In particular, at the center of the cylinder ($R=0$), the field has only the $z$ component. 
Using the formulas $\lim_{R \to 0}J_{0}(kR)=1$, $J_{1}(ka)=-a^{-1}\partial J_{0}(ka)/\partial k$, and $\int_{0}^{\infty} dk e^{-k|Z|} J_{0}(ka)=1/\sqrt{a^{2}+Z^{2}}$, 
we find from Eq. (\ref{eq:H_Z_Z}) that 
\begin{equation}
\begin{split}
  H_{ZZ}(0,0,Z)
  &=
  2\pi a M 
  \int_{0}^{\infty}
  dk 
  \left[
    {\rm sgn}(Z)
    e^{-k|Z|}
    -
    {\rm sgn}(Z+h)
    e^{-k|Z+h|}
  \right]
  J_{1}(ka)
\\
  &=
  -2\pi M 
  \int_{0}^{\infty}
  dk
  \left[
    {\rm sgn}(Z)
    e^{-k|Z|}
    -
    {\rm sgn}(Z+h)
    e^{-k|Z+h|}
  \right]
  \frac{\partial J_{0}(ka)}{\partial k}
\\
  &=
  2\pi M 
  \left[
    \frac{Z+h}{\sqrt{a^{2}+(Z+h)^{2}}}
    -
    \frac{Z}{\sqrt{a^{2}+Z^{2}}}
    +
    {\rm sgn}(Z)
    -
    {\rm sgn}(Z+h)
  \right]. 
\end{split}
\end{equation}
Here, we use the following partial integral formula, 
\begin{equation}
\begin{split}
&
  \int_{0}^{\infty}
  dk 
  \left[
    {\rm sgn}(Z)
    e^{-k|Z|}
    -
    {\rm sgn}(Z+h)
    e^{-k|Z+h|}
  \right]
  \frac{\partial J_{0}(ka)}{\partial k}
\\
  &=
  \int_{0}^{\infty}
  dk 
  \frac{\partial}{\partial k}
  \left\{
    \left[
      {\rm sgn}(Z)
      e^{-k|Z|}
      -
      {\rm sgn}(Z+h)
      e^{-k|Z+h|}
    \right]
    J_{0}(ka)
  \right\}
\\
  &+
  \int_{0}^{\infty}
  dk 
  \left[
    Z
    e^{-k|Z|}
    -
    (Z+h)
    e^{-k|Z+h|}
  \right]
  J_{0}(ka), 
\end{split}
\end{equation}
and use $\lim_{k \to \infty}J_{0}(ka)=0$. 
The term ${\rm sgn}(Z)-{\rm sgn}(Z+h)$ is finite ($-2$) only inside the ferromagnet ($-h<Z<0$), 
whereas it is zero outside the ferromagnet ($Z < -h$ or $Z>0$). 
This means that the field $H_{ZZ}$ are discontinuous on the surfaces located at $Z=0$ and $Z=-h$. 
This is because the magnetic poles appear on these surfaces in the present case. 




\begin{figure}
\centerline{\includegraphics[width=1.0\columnwidth]{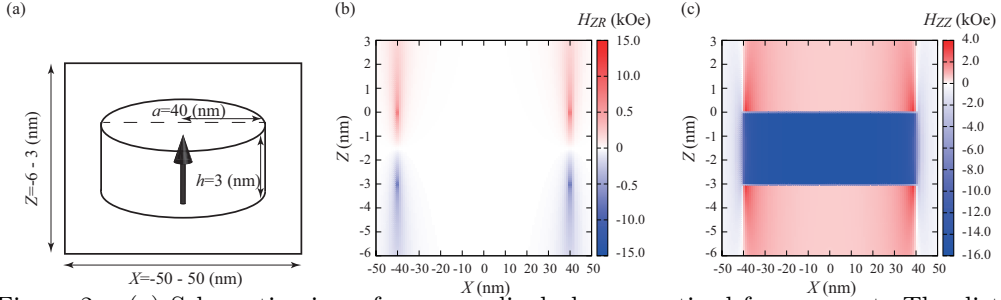}}\vspace{-3.0ex}
\caption{
         (a) Schematic view of a perpendicularly magnetized ferromagnet. 
             The distributions of $H_{ZR}$ and $H_{ZZ}$ on a plane at $Y=0$ nm for $-50 \le X \le 50$ nm and $-6 \le Z \le 3$ nm are shown in (b) and (c), respectively.  
         \vspace{-3ex}}
\label{fig:fig2}
\end{figure}




Figure \ref{fig:fig2} shows an example of the magnetic field distribution of a perpendicularly magnetized ferromagnet. 
The values of the parameters are derived from a recent experiment \cite{hiramatsu16}, 
where $M=1300$ emu/c.c., $a=40$ nm, and $h=3$ nm. 
As schematically shown in Fig. \ref{fig:fig2}(a), the field distribution is studied on a plane at $Y=0$ nm 
with the ranges of $-50 \le X \le 50$ nm and $-6 \le Z \le 3$ nm, 
whereas the ferromagnet is located at $-40 \le X \le 40$ nm and $-3 \le Z \le 0$ nm. 
Figures \ref{fig:fig3}(b) and \ref{fig:fig3}(c) show $H_{ZR}$ and $H_{ZZ}$, respectively. 
As shown, $H_{ZR}$ is dominated at the edges of the ferromagnet. 
On the other hand, $H_{ZZ}$ is discontinuous at the top and bottom surfaces of the ferromagnet, as mentioned above, 
whereas it is continuous at the other regions. 



\subsection{Demagnetization coefficient}

For the practical applications mentioned in Sec. \ref{sec:Introduction}, 
an evaluation of the magnetic field outside of ferromagnet is necessary 
because it determines the coupling strength of the ferromagnets. 
We note, however, that the formulas derived in the previous section are applicable to 
both inside and outside of the ferromagnet. 
In this section, we show that Eq. (\ref{eq:H_Z_Z}) can be used to evaluate the demagnetization coefficient of a cylinder ferromagnet. 


The analytical formula of the demagnetization coefficient can be derived by calculating the average of Eq. (\ref{eq:H_Z_Z}) over the volume of the ferromagnet. 
The volume integral of Eq. (\ref{eq:H_Z_Z}) is defined as 
\begin{equation}
\begin{split}
  I_{V}
  &=
  \int_{0}^{a}
  d R
  \int_{0}^{h}
  d Z^{\prime}
  \int_{0}^{2\pi}
  R d \theta
  H_{ZZ}
\\
  &=
  -2\pi a M 
  \int_{0}^{\infty} 
  dk 
  \int_{0}^{a}
  R d R 
  \int_{0}^{h}
  d Z^{\prime}
  \int_{0}^{2\pi}
  d \theta
  \left[
    e^{k(Z^{\prime}-h)}
    +
    e^{-kZ^{\prime}}
  \right]
  J_{1}(ka)
  J_{0}(kR), 
  \label{eq:demag_volume_integral}
\end{split}
\end{equation}
where $Z^{\prime}=Z+h$. 
The demagnetization coefficient in the $z$ direction, $N_{z}$, is defined as 
\begin{equation}
  4 \pi 
  N_{z} M
  =
  -\frac{I_{V}}{\pi a^{2}h}, 
\end{equation}
where $\pi a^{2}h$ is the volume of the ferromagnet. 
Using the formulas, 
\begin{equation}
  \int_{0}^{a}
  R d R 
  J_{0}(kR)
  =
  \frac{a J_{1}(ak)}{k},
\end{equation}
\begin{equation}
  \int_{0}^{\infty}
  dk 
  \frac{[J_{1}(ka)]^{2}}{k^{2}}
  =
  \frac{4a}{3\pi}, 
\end{equation}
\begin{equation}
  \int_{0}^{\infty}
  dk 
  e^{-kh}
  \frac{[J_{1}(ka)]^{2}}{k^{2}}
  =
  h
  \left[
    \frac{\sqrt{4a^{2}+h^{2}}}{6\pi a^{2}}
    h \mathsf{K}(\kappa)
    +
    \frac{(4a^{2}-h^{2}) \sqrt{4a^{2}+h^{2}}}{6\pi a^{2}h}
    \mathsf{E}(\kappa)
  \right]
  -
  \frac{h}{2},
\end{equation}
with $\kappa=2a/\sqrt{4a^{2}+h^{2}}$, 
the demagnetization coefficient $N_{z}$ is found to be 
\begin{equation}
\begin{split}
  N_{z}
  =
  \frac{1}{\tau}
  \left\{
    \frac{4}{3\pi}
    -
    \frac{4}{3\pi}
    \sqrt{1+\tau^{2}}
    \left[
      \tau^{2}
      \mathsf{K}
      \left(
        \frac{1}{\sqrt{1+\tau^{2}}}
      \right)
      +
      \left(
        1
        -
        \tau^{2}
      \right)
      \mathsf{E}
      \left(
        \frac{1}{\sqrt{1+\tau^{2}}}
      \right)
    \right]
    +
    \tau
  \right\}, 
  \label{eq:demag_coff_z}
\end{split}
\end{equation}
where $\tau=h/(2a)$. 

We note that Eq. (\ref{eq:demag_coff_z}) is identical to the formula derived in Refs. \cite{joseph66,tandon04}, 
where a different approach, based on the Fourier space formalism, to compute the demagnetization tensor was developed. 
Due to the axial symmetry, the demagnetization coefficients in the $x$ and $y$ directions are given by 
$N_{x}=N_{y}=(1-N_{z})/2$. 



\section{Magnetic field generated from an in-plane magnetized ferromagnet}
\label{sec:Magnetic field generated from an in-plane magnetized ferromagnet}

In this section, we calculate $\vec{H}_{X}$ corresponding to a magnetic field generated from an in-plane magnetized cylinder ferromagnet. 


\subsection{Magnetic potential}

We again use the cylindrical coordinate to derive the analytical formula of $\vec{H}_{X}$. 
The magnetic potential in this case is given by 
\begin{equation}
  V_{\rm i}
  =
  M 
  \int_{-h}^{0} 
  dz
  \int_{0}^{2\pi}
  a d \phi 
  \frac{\cos\phi}{\sqrt{(X-a \cos\phi)^{2}+(Y-a \sin\phi)^{2}+(Z-z)^{2}}}.
  \label{eq:V_i}
\end{equation}
We again introduce $R$, $\psi$, and $\theta$ as $X=R \cos\psi$, $Y=R \sin\psi$, and $\theta=\phi-\psi$. 
Using the Gegenbauer's addition theorem for the Bessel function,  
Eq. (\ref{eq:V_i}) can be rewritten as 
\begin{equation}
\begin{split}
  V_{\rm i}
  &=
  M
  \int_{-h}^{0}
  dz 
  \int_{-\psi}^{2\pi-\psi}
  a d \theta
  \int_{0}^{\infty}
  e^{-k|z-Z|}
  dk 
  \left[
    J_{0}(kR)
    J_{0}(ka)
    \cos(\theta+\psi)
  \right.
\\
  &\ \ \ \ \ \ 
  \left.
    +
    2 \sum_{m=1}^{\infty}
    J_{m}(kR)
    J_{m}(ka)
    \cos m \theta
    \cos(\theta+\psi)
  \right]
\\
  &=
  \begin{cases}
    2\pi a M 
    \cos\psi
    \int_{0}^{\infty}
    dk 
    \frac{{\rm sgn}(Z)e^{-k|Z|}-{\rm sgn}(Z+h)e^{-k|Z+h|}}{k}
    J_{1}(kR)
    J_{1}(ka)
    & 
    (Z<-h,\ Z>0) 
  \\
    2\pi a M 
    \cos\psi
    \int_{0}^{\infty}
    dk 
    \frac{2-e^{kZ}-e^{-k(Z+h)}}{k}
    J_{1}(kR)
    J_{1}(ka)
    &
    (-h<Z<0)
  \end{cases}.
  \label{eq:V_i_1}
\end{split}
\end{equation}
Here, we remind the readers that the regions $Z<-h$ and $Z>0$ correspond to outside the ferromagnet 
whereas the region $-h<Z<0$ corresponds to inside the ferromagnet. 
In the derivation, we use the following relations, 
\begin{equation}
  \int_{-h}^{0}
  dz
  \Theta(z-Z)
  e^{-k(z-Z)}
  =
  \begin{cases}
    \int_{-h}^{0} dz e^{-k(z-Z)} & (Z<-h) \\
    \int_{Z}^{0} dz e^{-k(z-Z)} & (-h<Z<0)
  \end{cases},
  \label{eq:integral_ex_1}
\end{equation}
\begin{equation}
  \int_{-h}^{0}
  dz
  \Theta(Z-z)
  e^{k(z-Z)}
  =
  \begin{cases}
    \int_{-h}^{0} dz e^{k(z-Z)} & (Z>0) \\
    \int_{-h}^{Z} dz e^{k(z-Z)} & (-h<Z<0)
  \end{cases}.
  \label{eq:integral_ex_2}
\end{equation}
Here the step function appears due to the following reason. 
The Gegenbauer's formula assumes that the exponent of the exponential is negative. 
Therefore, we use $e^{-k|z-Z|}$ in the first line of Eq. (\ref{eq:V_i_1}). 
Then, we use the step function to remove the symbol of the absolute value from this exponential. 
Physically, the use of the step function, as well as the different expressions of $V_{\rm i}$ for the inside and outside the ferromagnet, 
reflects the fact that a component ($H_{XX}$) of the magnetic field is discontinuous at the ferromagnetic interface, as shown below. 
We also use the following relation. 
\begin{equation}
\begin{split}
  &
  \int_{-\psi}^{2\pi-\psi}
  d \theta
  \sum_{m=1}^{\infty}
  J_{m}(kR)
  J_{m}(ka)
  \cos m\theta
  \cos(\theta+\psi)
\\
  &=
  \int_{-\psi}^{2\pi-\psi}
  d\theta
  J_{m}(kR)
  J_{m}(ka)
  \left\{
    \cos
    \left[
      \left(
        m + 1 
      \right)
      \theta
      +
      \psi
    \right]
    +
    \cos
    \left[
      \left(
        m
        -
        1
      \right)
      \theta
      -
      \psi
    \right]
  \right\}
\\
  &=
  J_{1}(kR)
  J_{1}(ka)
  \left[
    \frac{\sin(2\theta+\psi)}{2}
    +
    \theta
    \cos\psi
  \right]_{-\psi}^{2\pi-\psi}
\\
  &+
  \sum_{m=2}^{\infty}
  J_{m}(kR)
  J_{m}(ka)
  \left\{
    \frac{\sin[(m+1)\theta+\psi]}{m+1}
    +
    \frac{\sin[(m-1)\theta-\psi]}{m-1}
  \right\}_{-\psi}^{2\pi-\psi}
\\
  &=
  2\pi 
  J_{1}(kR)
  J_{1}(ka)
  \cos\psi 
\end{split}
\end{equation}



\subsection{Magnetic field in radial direction ($Z>0$ and $Z<-h$)}

The magnetic field in the radial direction for $Z>0$ and $Z<-h$ 
is given by $H_{XR}=-\partial V_{\rm i}/\partial R$, 
which can be expressed as 
\begin{equation}
\begin{split}
  H_{XR}
  &=
  -2\pi a M 
  \cos \psi 
  \int_{0}^{\infty}
  dk 
  \frac{{\rm sgn}(Z) e^{-k|Z|}- {\rm sgn}(Z+h) e^{-k|Z+h|}}{k}
  \frac{\partial J_{1}(kR)}{\partial R}
  J_{1}(ka)
\\
  &=
  -2\pi a M 
  \cos\psi 
  \left[
    \frac{{\rm sgn}(Z)U(Z) - {\rm sgn}(Z+h) U(Z+h)}{a}
  \right.
\\
  &\ \ \ \ 
  \left.
    -
    \frac{{\rm sgn}(Z) W(Z) - {\rm sgn}(Z+h) W(Z+h)}{R}
  \right],
  \label{eq:H_X_R}
\end{split}
\end{equation}
where we use the formula that the Bessel function satisfies, 
\begin{equation}
  \frac{\partial J_{1}(kR)}{\partial R}
  =
  k 
  J_{0}(kR)
  -
  \frac{J_{1}(kR)}{R}.
  \label{eq:Bessel_derivative_1}
\end{equation}
The function $U(Z)$ is defined by Eq. (\ref{eq:U_Z}). 
On the other hand, $W(Z)$ is given by 
\begin{equation}
\begin{split}
  W(Z)
  &=
  \int_{0}^{\infty}
  dk 
  e^{-k|Z|}
  \frac{J_{1}(kR) J_{1}(ka)}{k}
\\
  &=
  \frac{|Z| \mathsf{E}(\kappa_{1})}{\pi \kappa_{1} \sqrt{aR}}
  -
  \frac{|Z| \kappa_{1} (a^{2}+R^{2}+Z^{2}/2)}{2\pi (aR)^{3/2}}
  \mathsf{K}(\kappa_{1})
\\
  &+
  \frac{|a^{2}-R^{2}|}{2 \pi aR}
  \left[
    \mathsf{E}(\kappa_{1})
    \mathsf{F}(\beta_{1}^{\prime},\sqrt{1-\kappa_{1}^{2}})
    +
    \mathsf{K}(\kappa_{1})
    \mathsf{E}(\beta_{1}^{\prime},\sqrt{1-\kappa_{1}^{2}})
    -
    \mathsf{K}(\kappa_{1})
    \mathsf{F}(\beta_{1}^{\prime},\sqrt{1-\kappa_{1}^{2}})
  \right]
\\
  &+
  \frac{{\rm min}[a,R]}{2 {\rm max}[a,R]},
  \label{eq:W_Z}
\end{split}
\end{equation}
where $\kappa_{1}$ is given by Eq. (\ref{eq:kappa_1}), 
whereas $\beta_{1}^{\prime}$ is defined as 
\begin{equation}
  \beta_{1}^{\prime}
  =
  \sin^{-1}
  \frac{|Z|}{\sqrt{Z^{2}+(a-R)^{2}}}.
\end{equation}
We note that $W(Z+h)$ is obtained by replacing $\kappa_{1}$ and $\beta_{1}^{\prime}$ in $W(Z)$ with $\kappa_{2}$ and $\beta_{2}^{\prime}$, 
where $Z$ in $\kappa_{1}$ and $\beta_{1}^{\prime}$ are replaced with $Z+h$. 
We note that $H_{XR}(X,Y,Z_{0})=H_{XR}(X,Y,-Z_{0}-h)$ for a positive $Z_{0}$. 


\subsection{Magnetic field in radial direction ($-h<Z<0$)}

On the other hand, the magnetic field in the radial direction for $-h<Z<0$ is defined as 
\begin{equation}
  H_{XR}
  =
  -2\pi a M 
  \cos\psi
  \int_{0}^{\infty}
  dk 
  \frac{2 - e^{-k|Z|} - e^{-k|Z+h|}}{k}
  \frac{\partial J_{1}(kR)}{\partial R}
  J_{1}(ka). 
\end{equation}
Using Eq. (\ref{eq:Bessel_derivative_1}) and the integral formulas 
\begin{equation}
  \int_{0}^{\infty}
  dk 
  J_{0}(kR)
  J_{1}(ka)
  =
  \begin{cases}
    0 & (a<R) \\
    1/a & (a>R) \\
    1/(2a) & (a=R) 
  \end{cases}, 
\end{equation}
\begin{equation}
  \int_{0}^{\infty}
  dk 
  \frac{J_{1}(kR) J_{1}(ka)}{k}
  =
  \frac{{\rm min}[a,R]}{2 {\rm max}[a,R]},
  \label{eq:Bessel_integral_J1J1_k}
\end{equation}
we find that 
\begin{equation}
\begin{split}
  H_{XR}
  &=
  2\pi a M 
  \cos\psi
  \left[
    \frac{U(Z)+U(Z+h)}{a}
    -
    \frac{W(Z)+W(Z+h)}{R}
  \right]
\\
  &+
  \begin{cases}
    2 \pi (a/R)^{2} M \cos\psi & (a<R) \\
    -2\pi M \cos\psi & (a>R) \\
    0 & (a=R) 
  \end{cases}.
  \label{eq:H_X_R_inside}
\end{split}
\end{equation}
Since the magnetic pole appears on the lateral surface of the cylinder, 
$H_{XR}$ is discontinuous at $R=a$. 


On the other hand, we note that $H_{XR}$ is continuous on the top and bottom surfaces located at $Z=0$ and $Z=-h$. 
This fact can be confirmed as follows. 
Let us consider the surface at $Z=0$. 
Here, we should note the following relations, 
\begin{equation}
  \lim_{Z \to 0}
  \left[
    \frac{U(Z)}{a}
    -
    \frac{W(Z)}{R}
  \right]
  =
  \begin{cases}
    -a/(2R^{2}) & (a<R) \\
    1/(2a) & (a>R) \\
    0 & (a=R) 
  \end{cases}. 
\end{equation}
Then, both Eqs. (\ref{eq:H_X_R}) and (\ref{eq:H_X_R_inside}) become 
\begin{equation}
\begin{split}
  \lim_{Z \to 0}
  H_{XR}
  =&
  2\pi a M 
  \cos\psi
  \left[
    \frac{U(h)}{a}
    -
    \frac{W(h)}{R}
  \right]
\\
  &+
  \begin{cases}
    \pi M (a/R)^{2} \cos\psi & (a<R) \\
    -\pi M \cos\psi & (a>R) \\
    0 & (a=R)
  \end{cases}.
\end{split}
\end{equation}
We note that the same result can be obtained from Eq. (\ref{eq:H_X_R_inside}). 
The continuity of $H_{XR}$ at $Z=-h$ is also confirmed in a similar manner. 


\subsection{Magnetic field in perpendicular and circumferential directions ($Z>0$ and $Z<-h$)}

The magnetic field in the $z$ direction for $Z>0$ and $Z<-h$ 
is defined as $H_{XZ}=-\partial V_{\rm i}/\partial Z$, 
and is given by 
\begin{equation}
\begin{split}
  H_{XZ}
  &=
  2\pi a M 
  \cos\psi
  \int_{0}^{\infty}
  dk 
  \left(
    e^{-k|Z|}
    -
    e^{-k|Z+h|}
  \right)
  J_{1}(kR)
  J_{1}(ka)
\\
  &=
  4 \sqrt{\frac{a}{R}}
  M 
  \cos\psi
  \left\{
    \frac{1}{\kappa_{1}}
    \left[
      \left(
        1
        -
        \frac{\kappa_{1}^{2}}{2}
      \right)
      \mathsf{K}(\kappa_{1})
      -
      \mathsf{E}(\kappa_{1})
    \right]
    -
    \frac{1}{\kappa_{2}}
    \left[
      \left(
        1
        -
        \frac{\kappa_{2}^{2}}{2}
      \right)
      \mathsf{K}(\kappa_{2})
      -
      \mathsf{E}(\kappa_{2})
    \right]
  \right\}. 
  \label{eq:H_X_Z}
\end{split}
\end{equation}
We note that Eq. (\ref{eq:H_X_Z}) is identical to Eq. (\ref{eq:H_Z_R}) except the factor $\cos\psi$. 
We also note that $H_{XZ}(X,Y,Z_{0})=-H_{XZ}(X,Y,-Z_{0}-h)$ for a positive $Z_{0}$. 

On the other hand, the field in the circumferential direction for $Z>0$ and $Z<-h$ 
is defined as $H_{X \psi}=-R^{-1}\partial V_{\rm i}/\partial \psi$, 
and is given by 
\begin{equation}
\begin{split}
  H_{X\psi}
  &=
  \frac{2\pi aM \sin \psi}{R}
  \int_{0}^{\infty}
  dk
  \frac{{\rm sgn}(Z)e^{-k|Z|}-{\rm sgn}(Z+h)e^{-k|Z+h|}}{k}
  J_{1}(kR)
  J_{1}(ka)
\\
  &=
  2 \pi a M 
  \sin\psi
  \frac{{\rm sgn}(Z)W(Z) - {\rm sgn}(Z+h)W(Z+h)}{R},
  \label{eq:H_X_psi}
\end{split}
\end{equation}
where $W(Z)$ is defined by Eq. (\ref{eq:W_Z}). 
We note that $H_{X\psi}(X,Y,Z_{0})=H_{X\psi}(X,Y,-Z_{0}-h)$ for a positive $Z_{0}$. 


\subsection{Magnetic field in perpendicular and circumferential directions ($-h<Z<0$)}

The magnetic field in the $z$ direction for $-h<Z<0$ is given by
\begin{equation}
  H_{XZ}
  =
  2\pi a M 
  \cos \psi 
  \int_{0}^{\infty}
  dk 
  \left(
    e^{-k|Z|}
    -
    e^{-k|Z+h|}
  \right)
  J_{1}(kR)
  J_{1}(ka),
  \label{eq:H_X_Z_inside}
\end{equation}
which is identical to Eq. (\ref{eq:H_X_Z}). 
Therefore, $H_{XZ}$ can be expressed by the same equation, Eq. (\ref{eq:H_X_Z}) for an arbitrary $Z$. 

On the other hand, the field in the circumferential direction for $-h<Z<0$ is 
\begin{equation}
\begin{split}
  H_{X\psi}
  &=
  \frac{2\pi aM \sin\psi}{R}
  \int_{0}^{\infty}
  dk 
  \frac{2-e^{-k|Z|}-e^{-k|Z+h|}}{k}
  J_{1}(kR)
  J_{1}(ka)
\\
  &=
  \frac{2\pi aM}{R} 
  \frac{{\rm min}[a,R]}{{\rm max}[a,R]}
  \sin \psi
  -
  \frac{2\pi aM \sin\psi}{R}
  \left[
    W(Z)
    +
    W(Z+h)
  \right], 
  \label{eq:H_X_psi_inside}
\end{split}
\end{equation}
where we use Eq. (\ref{eq:Bessel_integral_J1J1_k}). 
We can again confirm that $H_{X\psi}$ is continuous on the ferromagnetic top and bottom surfaces. 
For example, at $Z=0$, 
using $\lim_{Z \to 0}W(Z)={\rm min}[a,R]/(2 {\rm max}[a,R])$, 
both Eqs. (\ref{eq:H_X_psi}) and (\ref{eq:H_X_psi_inside}) give 
\begin{equation}
  \lim_{Z \to 0}
  H_{X \psi}
  =
  \frac{2 \pi a M}{R} 
  \sin \psi
  \left[
    \frac{{\rm min}[a,R]}{2 {\rm max}[a,R]}
    -
    W(h)
  \right].
\end{equation}


\subsection{Short summary of this section}

Equations (\ref{eq:H_X_R}), (\ref{eq:H_X_R_inside}), (\ref{eq:H_X_Z}), (\ref{eq:H_X_Z_inside}), (\ref{eq:H_X_psi}), and (\ref{eq:H_X_psi_inside}) 
give analytical formulas of the magnetic field generating from an in-plane magnetized ferromagnet, 
and are the main results in this section. 
The magnetic field is continuous at the top and bottom surfaces at $Z=0$ and $Z=-h$. 
On the lateral surface $R=a$, on the other hand, $H_{XR}$ is discontinuous due to the magnetic pole. 


In particular, at the center of the cylinder ($R=0$), 
the fields in the radius and circumferential directions are finite and are given by 
\begin{equation}
\begin{split}
  \lim_{R \to 0}
  H_{XR}
  &=
  -\pi a M 
  \cos\psi
  \int_{0}^{\infty}
  dk 
  \left[
    {\rm sgn}(Z)
    e^{-k|Z|}
    -
    {\rm sgn}(Z)
    e^{-k|Z+h|}
  \right]
  J_{1}(ka)
\\
  &=
  -\pi M 
  \cos\psi
  \left[
    \frac{Z+h}{\sqrt{a^{2}+(Z+h)^{2}}}
    -
    \frac{Z}{\sqrt{a^{2}+Z^{2}}}
  \right], 
\end{split}
\end{equation}
\begin{equation}
\begin{split}
  \lim_{R \to 0}
  H_{X \psi}
  &=
  \pi a M 
  \sin\psi
  \int_{0}^{\infty}
  dk 
  \left[
    {\rm sgn}(Z)
    e^{-k|Z|}
    -
    {\rm sgn}(Z+h)
    e^{-k|Z+h|}
  \right]
  J_{1}(ka)
\\
  &=
  \pi M 
  \sin\psi
  \left[
    \frac{Z+h}{\sqrt{a^{2}+(Z+h)^{2}}}
    -
    \frac{Z}{\sqrt{a^{2}+Z^{2}}}
  \right], 
\end{split}
\end{equation}
where we use 
\begin{equation}
  \lim_{R \to 0}
  \frac{\partial J_{1}(kR)}{\partial R}
  =
  \lim_{R \to 0}
  \left[
    k 
    J_{0}(kR)
    -
    \frac{J_{1}(kR)}{R}
  \right]
  =
  \frac{k}{2}.
\end{equation}
In the Cartesian coordinate, only the $x$ component is finite at the center, 
whereas outside the ferromagnet is given by 
\begin{equation}
  H_{XX}(0,0,Z)
  =
  -\pi M 
  \left[
    \frac{Z+h}{\sqrt{a^{2}+(Z+h)^{2}}}
    -
    \frac{Z}{\sqrt{a^{2}+Z^{2}}}
  \right]. 
\end{equation}




\begin{figure}
\centerline{\includegraphics[width=1.0\columnwidth]{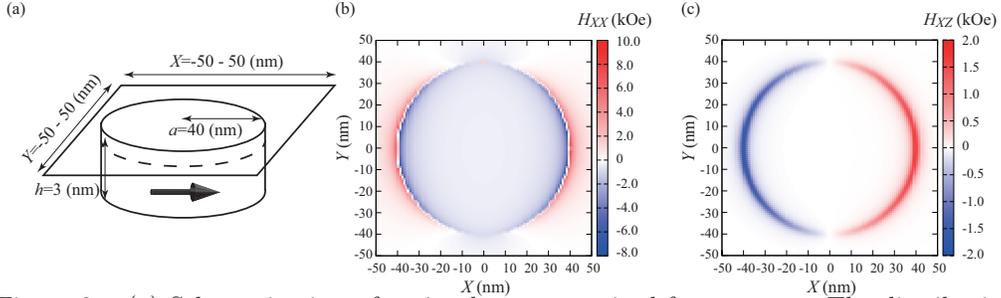}}\vspace{-3.0ex}
\caption{
         (a) Schematic view of an in-plane magnetized ferromagnet. 
             The distributions of $H_{XX}$ and $H_{XZ}$ on a plane at $Z=-1$ nm for $-50 \le X \le 50$ nm and $-50 \le Y \le 50$ nm are shown in (b) and (c), respectively.  
         \vspace{-3ex}}
\label{fig:fig3}
\end{figure}




Figure \ref{fig:fig3} shows an example of the magnetic field distribution of an in-plane magnetized ferromagnet. 
The values of the parameters are same with those used in Fig. \ref{fig:fig2}. 
As schematically shown in Fig. \ref{fig:fig3}(a), the field distribution is studied on a plane at $Z=-1$ nm 
with the ranges of $-50 \le X \le 50$ nm and $-50 \le Y \le 50$ nm. 
Figures \ref{fig:fig3}(b) and \ref{fig:fig3}(c) show $H_{XX}=H_{XR}\cos\psi - H_{X\psi}\sin\psi$ and $H_{XZ}$, respectively. 
As shown, $H_{XX}$ is discontinuous at the lateral surface ($X^{2}+Y^{2}=a^{2}$) due to the magnetic poles on the surface. 
On the other hand, $H_{ZZ}$ is continuous, and is positive (negative) for a positive (negative) $X$. 



\section{Application}
\label{sec:Application}


A direct application of the above results is an evaluation of the magnetic field generated by a spin torque oscillator. 
A spin torque oscillator is a microwave generator consisting of a nanostructured ferromagnetic/nonmagnetic multilayer. 
Usually, it consists of two ferromagnets named free and pinned layer. 
Injecting an electric current into the structure, the pinned layer polarizes the spin of the conducting electrons. 
The spin transfer from the conducting electrons to the local magnetization in the free layer excites spin transfer torque \cite{slonczewski96}, 
resulting in an excitation of an auto-oscillation of the magnetization \cite{kubota13,taniguchi14}. 
It has been proposed in the field of magnetic recording that 
the oscillating magnetic field emitted from the spin torque oscillator is useful to reduce 
the magnitude of the writing magnetic field of high-density magnetic recording media. 
The method of this magnetization switching is called microwave assisted magnetization reversal (MAMR) or microwave assisted magnetization switching (MAS) 
\cite{zhu08,kudo14,suto14,kudo15,suto16PRA,suto16,okamoto12,tanaka13,okamoto15,taniguchi15a,taniguchi15b,suto17}. 
The magnitude of the magnetic field emitted from the spin torque oscillator determines 
the writing magnetic field in MAMR \cite{taniguchi14PRB,taniguchi16PRB}. 
The magnitude of the magnetic field from the spin torque oscillator has been not, however, fully studied yet 
both experimentally and theoretically. 
In the following, let us provide the evaluation method of such magnetic field. 

\begin{figure}
\centerline{\includegraphics[width=1.0\columnwidth]{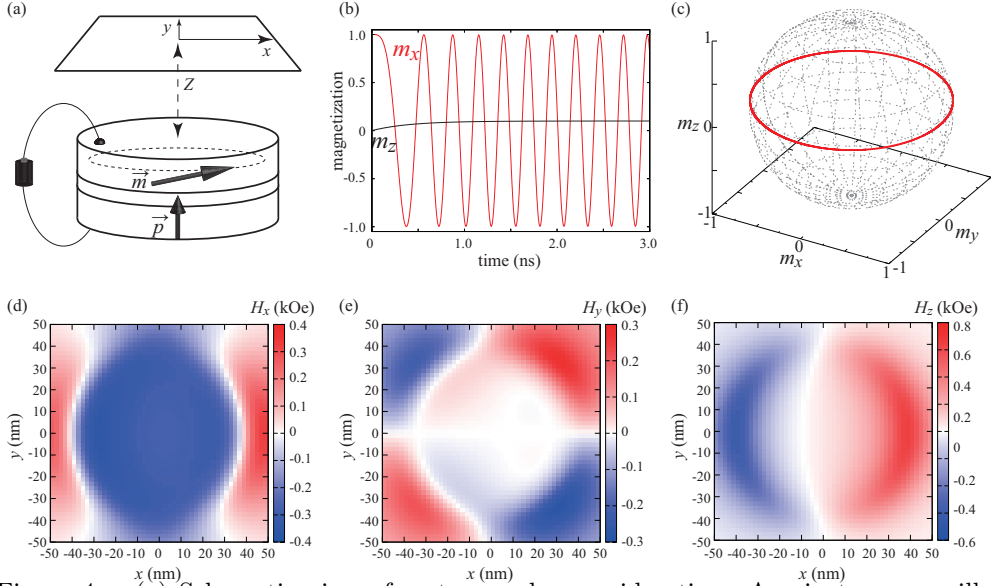}}\vspace{-3.0ex}
\caption{
         (a) Schematic view of system under consideration. 
             A spin torque oscillator consisting of an in-plane magnetized free layer and a perpendicularly magnetized pinned layer shows an auto-oscillation 
             of the magnetization $\vec{m}$ in the free layer due to the current injection. 
             The distribution of the magnetic field in the $xy$ plane from the top surface of the free layer within a distance of $Z=10$ nm is calculated. 
         (b) The evolutions of $x$ (red) and $z$ (black) component of $\vec{m}$ by the current injection. 
             The initial state is set to be $\vec{m}(0)=(1,0,0)$. 
         (c) Schematic illustration of a steady state auto-oscillation in the free layer. 
         (d)-(f) The $x$, $y$, and $z$ components of the magnetic field in the $xy$ plane. 
             The radius of the spin torque oscillator is set to be 40 nm. 
         \vspace{-3ex}}
\label{fig:fig4}
\end{figure}

A potential structure of a spin torque oscillator for MAMR is 
a ferromagnetic multilayer consisting of an in-plane magnetized free layer and a perpendicularly magnetized pinned layer 
\cite{suto12,lee05,firastrau07,ebels08,silva10,igarashi10,taniguchi16,taniguchi16APEX}, 
which is schematically shown in Fig. \ref{fig:fig4}(a). 
We denote the unit vectors pointing in the directions of the magnetizations 
in the free and pinned layers as $\vec{m}$ and $\vec{p}$, respectively. 
The magnetization in the pinned layer points to the perpendicular ($z$) direction, i.e., $\vec{p}=\vec{e}_{z}$. 
The positive electric current corresponds to an electron flowing from the free to pinned layer, 
i.e., the spin transfer torque excited on the free layer by a positive current prefers 
the antiparallel alignment of the magnetizations $\vec{m}$ and $\vec{p}$. 
The magnetization dynamics in the free layer is described by the Landau-Lifshitz-Gilbert (LLG) equation given by 
\begin{equation}
  \frac{d \vec{m}}{d t}
  =
  -\gamma
  \vec{m}
  \times
  \vec{H}_{m}
  -
  \gamma
  H_{\rm s}
  \vec{m}
  \times
  \left(
    \vec{p}
    \times
    \vec{m}
  \right)
  +
  \alpha
  \vec{m}
  \times
  \frac{d \vec{m}}{dt},
  \label{eq:LLG}
\end{equation}
where $\gamma$ and $\alpha$ are the gyromagnetic ratio and the Gilbert damping constant, respectively. 
The magnetic field acting on the magnetization in the free layer is 
\begin{equation}
  \vec{H}_{m}
  =
  -4 \pi M 
  \tilde{N}_{z}
  m_{z}
  \vec{e}_{z}.
  \label{eq:field}
\end{equation}
where a stray field from the pinned layer is neglected because the pinned layer usually consists of anti-ferromagnetically coupled two ferromagnets \cite{hiramatsu16}, 
and thus, the stray field might be totally small. 
The demagnetization coefficient $\tilde{N}_{z}=N_{z}-N_{x}$ is estimated to be 0.85 by using the parameters listed below. 
The strength of the spin transfer torque, $H_{\rm s}$, is given by 
\begin{equation}
  H_{\rm s}
  =
  \frac{\hbar \eta J}{2eMd},
\end{equation}
where $\eta$, $J$, and $d$ are the spin polarization, electric current density, 
and thickness of the free layer, respectively, 
whereas $\hbar$ and $e=|e|$ are the reduced Planck constant and elementary charge, respectively. 
The spin torque asymmetry \cite{taniguchi16APEX} is neglected, for simplicity. 


The spin transfer torque by a positive (negative) current moves the magnetization to the negative (positive) $z$ direction. 
Then, a stable oscillation of $\vec{m}$ with a constant tilted $\theta$ angle from the $z$ axis is excited, 
where $m_{z}^{*} \equiv \cos\theta$ is given by 
\begin{equation}
  m_{z}^{*}
  =
  \frac{-H_{\rm s}/\alpha}{4\pi M \tilde{N}_{z}}. 
\end{equation}
Here, we assume that $|H_{\rm s}/(\alpha 4\pi M \tilde{N}_{z})|<1$. 
If this condition is not satisfied, the spin torque overcomes the damping torque due to the demagnetization field, 
and the magnetization becomes parallel to the $z$ axis. 
In the following, we consider the magnetization dynamics excited by a negative current for convention. 
In this case, the magnetization $\vec{m}$ moves to the positive $z$ direction, as schematically shown in Fig. \ref{fig:fig1}(a). 
The values of the parameters used in the following calculation are taken from typical experiment \cite{hiramatsu16}, 
where $M=1300$ emu/c.c., $\gamma=1.764 \times 10^{7}$ rad/(Oe s), $d=3$ nm, $\eta=0.5$, and $\alpha=0.01$. 
Figure \ref{fig:fig4}(b) shows an example of the time evolution of the $x$ and $z$ components of $\vec{m}$ 
obtained by solving Eq. (\ref{eq:LLG}) numerically. 
The current density is set as $J \simeq -3.3 \times 10^{6}$ A/cm${}^{2}$ so that $m_{z}^{*}=0.1$. 
Starting from the initial state located in the film plane ($m_{z}=0$), which corresponds to an energetically stable state, 
the spin torque moves the magnetization to the positive $z$ direction. 
When $m_{z}$ reaches $m_{z}^{*}$, a steady precession around $z$ axis is stabilized. 
Figure \ref{fig:fig4}(c) shows the precession trajectory in this steady state.




Let us evaluate the magnetic field emitted from this type of the spin torque oscillator. 
We focus on the distribution of the magnetic field generated above the free layer's top surface 
within a distance of $Z=10$ nm \cite{suto14}, as schematically shown in Fig. \ref{fig:fig4}(a). 
The magnetic field is calculated when the magnetization comes to a position at $\vec{m}=(\sqrt{1-m_{z}^{*2}},0,m_{z}^{*})$. 
The radius of the spin torque oscillator is $40$ nm \cite{hiramatsu16}, 
whereas the field distribution is studied for the region of $-50 \le x \le 50$ nm and $-50 \le y \le 50$ nm. 
Figures \ref{fig:fig4}(d)-(f) show the magnetic field magnitudes of the $x$, $y$, and $z$ components, respectively. 
The field magnitude is estimated to be on the order of 100 Oe. 
Also, it can be seen that the field is finite in the region just above the spin torque oscillator, i.e., $x^{2}+y^{2} \le 40^{2}$ nm${}^{2}$, 
whereas at the outside of this region the field is relatively small. 


The evaluation of the magnetic field generated by the spin torque oscillator will be useful for the device design of the magnetic recording based on MAMR. 
In MAMR, the reduction of the writing field strongly depends on the frequency and amplitude of a microwave field \cite{okamoto15}. 
For example, the writing field is minimized when the microwave frequency is given by \cite{taniguchi14PRB,taniguchi16PRB} 
\begin{equation}
  f_{\rm MAMR} 
  =
  \frac{\gamma H_{\rm K}}{2\pi}
  \frac{(H_{\rm ac}/H_{\rm K})^{2/3}}{\sqrt{1-(H_{\rm ac}/H_{\rm K})^{2/3}}}
  \left[
    2
    -
    \frac{5}{3}
    \left(
      \frac{H_{\rm ac}}{H_{\rm K}}
    \right)^{2/3}
  \right],
  \label{eq:critical_frequency}
\end{equation}
where $H_{\rm ac}$ is the amplitude of the microwave field whereas $H_{\rm K}$ is the uniaxial anisotropy field in the recording media. 
In recent experiments, Oersted field generated from an electric current flowing above a recording bit is used as a microwave field \cite{suto17}. 
For MAMR, however, a spin torque oscillator is expected to act as a microwave source \cite{zhu08}. 
In this case, the magnitude of the microwave field can be evaluated by a similar manner with Figs. \ref{fig:fig4}(d)-(f), 
while its frequency is given by 
\begin{equation}
\begin{split}
  f
  &=
  \frac{\gamma}{2\pi}
  4\pi M 
  \tilde{N}_{z} 
  |m_{z}^{*}|
\\
  &=
  \frac{\gamma}{2\pi}
  \frac{|H_{\rm s}|}{\alpha}. 
  \label{eq:frequency_STO}
\end{split}
\end{equation}
Therefore, for example, the writing field of MAMR will be optimized by designing the material parameters of the spin torque oscillator and magnetic recording bit 
to make the magnetic field generated by the spin torque oscillator, as shown in Fig. \ref{fig:fig4}, and the oscillation frequency given by Eq. (\ref{eq:frequency_STO}) 
satisfying Eq. (\ref{eq:critical_frequency}). 
For example, it is estimated from Fig. \ref{fig:fig4} that the microwave field $H_{\rm ac}$, which corresponds to $H_{x}$ in Fig. \ref{fig:fig4}(d), 
emitted from the spin torque oscillator in the present calculation is 400 Oe at maximum. 
Also, the oscillation frequency of $\vec{m}$ is $f \simeq 3.9$ GHz. 
Then, $H_{\rm K}$ satisfying $f=f_{\rm MAMR}$ is estimated as 2.8 kOe. 


In the above calculation, the anisotropy field of a recording bit is estimated from the properties of the spin torque oscillator, 
such as the oscillation frequency and amplitude, to satisfy the opitimum condition of MAMR. 
However, usually, the value of the anisotropy field in the recording bit is determined to guarantee high thermal stability. 
Therefore, in reality, the present calculation will be used to design the properties of the spin torque oscillator, 
such as the saturation magnetization and damping constant, 
to satisfy the optimized condition $f=f_{\rm MAMR}$ as much as possible. 


\section{Summary}
\label{sec:Summary}

In summary, an analytical formulation computing a magnetic field generated 
from an uniformly magnetized cylinder ferromagnet was developed. 
The magnetic field is defined by Eq. (\ref{eq:field_def}), 
where $\vec{H}_{Z(X)}$ is the field generated from a perpendicularly (an in-plane) magnetized ferromagnet, 
whereas $\theta$ is the tilted angle of the magnetization from the perpendicular ($z$) axis. 
The analytical formulas of the components of $\vec{H}_{Z}$ are Eqs. (\ref{eq:H_Z_R}) and (\ref{eq:H_Z_Z}).
On the other hand, the components of $\vec{H}_{X}$ are given by Eqs. (\ref{eq:H_X_R}), (\ref{eq:H_X_Z}), and (\ref{eq:H_X_psi}) for $Z>0$ and $Z<-h$, 
whereas those for $-h<Z<0$ are given by Eqs. (\ref{eq:H_X_R_inside}), (\ref{eq:H_X_Z_inside}), and (\ref{eq:H_X_psi_inside}). 
The comprehensive description clearly shows the discontinuity of the magnetic field at surfaces due to magnetic pole. 
The validity of the derived formulas was confirmed by calculating the demagnetization field 
and comparing it with the result in earlier studies. 
The magnetic field generated from a spin torque oscillator is also studied, 
which will be useful for the device designing of the magnetic recording based on microwave assisted magnetization reversal. 



\section*{Acknowledgment}
The author express the deepest gratitude to Takehiko Yorozu for his great contribution to this work. 









\end{document}